# Room temperature ferromagnetism in carbon-implanted ZnO

Shengqiang Zhou[1], Qingyu Xu[1,2], Kay Potzger[1], Georg Talut[1], Rainer Grötzschel[1],

Jürgen Fassbender[1], Mykola Vinnichenko[1], Jörg Grenzer[1], Manfred Helm[1], Holger

Hochmuth[3], Michael Lorenz[3], Marius Grundmann[3], Heidemarie Schmidt[1]

[1]Institut für Ionenstrahlphysik und Materialforschung, Forschungszentrum Dresden-
Rossendorf e.V., Bautzner Landstraße 128, 01328 Dresden, Germany

[2]Department of Physics, Southeastern University, Nanjing 211189, China

[3]Institut für Experimentelle Physik II, Universität Leipzig, Linnéstraße 5, 04103 Leipzig,
Germany

Abstract:

Unexpected ferromagnetism has been observed in carbon doped ZnO films grown by pulsed laser deposition [Phys. Rev. Lett. **99**, 127201 (2007)]. In this letter, we introduce carbon into ZnO films by ion implantation. Room temperature ferromagnetism has been observed. Our analysis demonstrates that (1) C-doped ferromagnetic ZnO can be achieved by an alternative method, i.e. ion implantation, and (2) the chemical involvement of carbon in the ferromagnetism is indirectly proven.

The practical application of diluted magnetic semiconductors (DMS) in spintronics requires the DMS to exhibit ferromagnetism above room temperature. After the prediction that a Curie temperature higher than room temperature can be achieved by transition metal (TM) doping, ZnO has been intensively studied [1, 2]. Room temperature ferromagnetism has been reported by many researchers for 3$d$ TM doped ZnO [3-5]. However, the observed ferromagnetism is always rather weak and the detected magnetization might arise from ferromagnetic impurities [6, 7]. Another source for weak ferromagnetic signals are intrinsic defect complexes which occur in ZnO of low crystalline quality, e.g. if grown under $N_2$ atmosphere [8] or if vacuum annealed [9]. Such ferromagnetic defects also can be produced by Fe or Ar ion beams [9, 10].

Despite such drawbacks, H. Pan *et al.* reported strong room temperature ferromagnetism in C-doped ZnO films grown by pulsed laser deposition [11]. Together with the first-principles calculations, evidence is given that carbon ions substitute for oxygen and their p-orbits contribute the local moments. Since carbon is, ca. 5 times lighter than transition metal, it can be incorporated into ZnO by ion implantation while much less defects are created [12]. Recently, the equal-term applicability of thin-film growth and carbon ion implantation has been demonstrated for the ferromagnetic semiconductor C:$Mn_5Si_3$ [13]. We utilized various techniques to characterize the structural and magnetic properties of the C implanted ZnO films with different C-concentrations, as well as of a Ne-implanted ZnO and a C-implanted Ge reference samples. The aim is to check whether ferromagnetic ZnO can be realized by C-implantation, and whether carbon, or indeed defects play the active role in generating the observed ferromagnetism.

260 nm thick ZnO films were grown from a ZnO ceramic target on 10×10 mm$^2$ *a*-plane sapphire substrates by pulsed laser deposition (PLD) using a KrF excimer laser in $O_2$ atmosphere (0.2 Pa) with a substrate temperature $T_s$ of 715 $^o$C and subsequently implanted.

The TRIM code was used to simulate the depth dependent distribution of implanted C in the ZnO films [14]. In order to obtain a box-like C distribution in the ZnO films, four different energies with different fluences of C were implanted in the same film at room temperature. For instance, for 1 *at.*% C-doped ZnO, 70 keV, 35 keV, 17.5 keV and 8.8 keV with a fluence of $8.2\times10^{15}$, $3.28\times10^{15}$, $9.84\times10^{14}$, and $9.84\times10^{14}$ cm$^{-2}$ were applied, respectively. As comparison, Ne ions were implanted with the same energy choice but with a fluence of $16.4\times10^{15}$, $6.56\times10^{15}$, $1.97\times10^{15}$, and $1.97\times10^{15}$ cm$^{-2}$. Additionally, in order to exclude the possible contamination of carbon beam a Ge wafer was implanted with carbon at the same implanter with the ion energy of 70 keV and 35 keV, the fluences of $4.1\times10^{16}$ and $1.6\times10^{16}$ cm$^{-2}$, respectively. Compared with the C-fluence series, the middle fluence for Ne was chosen to produce mild amount of defects, while the largest C-fluence for the Ge sample was used to receive the largest amount of contamination if there was. The magnetization *M* versus temperature *T* from 5 to 300 K and versus magnetic field *H* was measured with a superconducting quantum interference device magnetometer (SQUID) with the magnetic field applied in the film plane. All samples were investigated using Rutherford backscattering/channelling spectrometry (RBS/C) to check the defects induced by ion implantation. The crystal lattice disorder upon implantation is quantified by the parameter $\chi_{min}$, which is the ratio of the backscattering yield at channelling condition to that at random beam incidence. Here $\chi_{min}$ is given for the Zn signal. The crystal structure of the films was probed by x-ray diffraction (XRD) *θ-2θ* scans using a Cu $K_\alpha$ source.

Magnetic properties of the pure ZnO, C- and Ne-implanted ZnO films as well as a C-implanted Ge wafer were investigated. Fig. 1 shows the *M-H* curves at 5 K of virgin ZnO, Ne-implanted, and 5 *at.*% C-implanted ZnO films with the area of one cm$^2$ at 5 K. For the virgin and Ne implanted ZnO film, only a linear diamagnetic signal is observed at 5 K. After implantation with 5 *at.*% C, an overlap of M-H hysteresis and the diamagnetic signal already probed on unimplanted ZnO is observed at 5 K. The inset of Fig. 1 displays the M-H curve measured at 4 K for C-implanted Ge, which exhibits only diamagnetism and no difference compared with the virgin Ge. This observation confirms the non-contamination of the carbon-beam by *3d* transition metals.

We subtracted the diamagnetic contribution determined from the high-field linear part of *M-H* loops to get the ferromagnetic magnetization of the C-implanted films. Fig. 2(a) shows the *M-H* curves for two selected samples, 1 *at*.% C and 5 *at*.% C implanted ZnO films, at 5 K and 300 K. A clear hysteresis loop can be observed, indicating the ferromagnetic properties. The concentration-dependent saturation magnetization $M_s$ has been plotted in Fig. 3(d) and (e) being normalized by the carbon concentration in units of $\mu_B$/C or being normalized in units of $10^{-5}$ emu per sample with the area of one cm$^2$. The saturation magnetization of 1 *at*.% C-implanted ZnO is about 0.24 $\mu_B$/C, and decreases to about 0.06 $\mu_B$/C for 5 *at*.% C-implanted ZnO. The hysteresis loop can also be observed at 300 K, as can be seen exemplarily in the *M-H* curve for 5 *at*.% C-implanted ZnO in the inset of Fig. 2(a), indicating the Curie temperature is above room temperature. Fig. 2(b) shows the field cooled (FC) and zero field cooled (ZFC) curve for 5 *at*.% implanted ZnO measured in an applied field of 100 Oe. Similar FC and ZFC curves were also observed in x *at*.% C-implanted ZnO (x=1,2,3,4). The separation between the FC and ZFC curves further demonstrates the ferromagnetic properties of C-implanted ZnO. It must be noted that the FC curve was measured after the ZFC curve, the magnetization was measured with field cooled from 300 K to 5 K under 100 Oe. Thus the separation between ZFC and FC disappears at 300 K. The real Curie temperature can be assumed to be higher than room temperature, which is also supported by the slight decrease of saturation magnetization at 300 K as shown in the inset of Fig. 2(a). In contrast, for the virgin and Ne implanted ZnO film, only a linear diamagnetic signal is found at 5 K (Fig. 1). Thus, purely defect induced ferromagnetism can be excluded. Further support for the exclusion of ferromagnetic defects can be found by annealing. The 4 *at.* % C-implanted sample was annealed in air for 2 hours at 400 $^{o}$C [10]. After annealing, the sample magnetization is not significantly changed, but the coercivity is increased from ~100 Oe to ~250 Oe. Thus, we can exclude the purely defect-induced ferromagnetism, since in such case the saturation magnetization should significantly drop with post-annealing [10].

No impurity phases were observed in the *θ-2θ* XRD patterns of the unimplanted and C-implanted wurtzite ZnO (not shown). However, besides the (002) peak of ZnO, the (100) peak can also be observed in virgin ZnO films. This indicates the coexistence of (001) and (100) textured grains in the films. After C implantation, the (100) peak position shifts significantly to smaller angles. The lattice constants calculated from the position of ZnO (100) and (002) peaks are summarized in Fig. 3(a) and (b). The *a*-lattice constant of the (100) oriented grains gradually increases with increasing C concentration, while the *c*-lattice constant of the (001) oriented grains only slightly increases after C implantation and reaches an average value of 0.525 nm at larger C concentrations. As the ionic radius of $C^{4-}$ (0.260 nm) is much larger than that of $O^{2-}$ (0.140 nm) [15], the substitution of O by C will expand the lattice. However, the expansion of the ZnO lattice might also originate from C atoms located at interstitial sites, or from defects induced by ion implantation. In Fig. 3(c), the minimum yield of RBS/C, $\chi_{min}$ of the Zn signal, is shown as a function of carbon concentration. For the virgin sample, $\chi_{min}$ is around 26%, which is acceptable considering the two kinds of ZnO crystallite orientation. With increasing implantation fluences, $\chi_{min}$ gradually increases, indicating more defects generated inside the ZnO matrix. However, ZnO is an irradiation-hard material and still shows the channelling effect ($\chi_{min}$ of 73%) after the largest fluence implantation. In order to understand the observed ferromagnetism, all the measured structural and magnetic parameters are compared in Fig. 3. There is a scaling effect between lattice expansion and RBS channelling $\chi_{min}$: both parameters increase gradually with increasing carbon concentration.

This proves that in the region of larger fluences, lattice expansion is induced rather by defects than by carbon substitution. For ferromagnetism induced by substitutional carbon the maximum magnetization thus is expected for fluences/concentrations where its crystalline neighbourhood is still intact, i.e. below disorder saturation at 3% carbon concentration [Fig. 3(c)]. Exactly this is what we observe, i.e. both the normalized saturation magnetizations [Fig. 3 (d) and (e)] reach their maximum value for 1 and 2 at% carbon, respectively. It must be noted that the saturation magnetic moment per carbon (highest value of 0.24 $\mu_B$/C for 1 *at.*% C-implanted ZnO at 5 K) is much smaller than the theoretically predicted value of 2.02 $\mu_B$/C and the experimentally reported value of 2.5 $\mu_B$/C [11]. We relate this behaviour to non-uniform distribution of the C ions inside the ZnO host lattice and thus a large number of magnetically inactive C-ions. The origin of the ferromagnetic signal must be related to the chemical presence of C [11] and not to implantation induced defects, since the Ne implanted sample is diamagnetic.

A measurement confirming the substitution of carbon is surely helpful to understand the ferromagnetic coupling mechanism. Channelling RBS and particle induced X-ray emission can probe the lattice location of heavy elements [16, 17], while channelling nuclear reaction analysis (NRA) would be suitable for the lattice location of light element, e.g. C [18]. We performed channelling NRA using the $^{12}$C(d, p)$^{13}$C reaction. Implanted carbon can be easily detected, but no channelling effect can be found. This is due to the fact that the applicability of channelling techniques depends on the crystalline quality. For our C-implanted ZnO, the $\chi_{min}$ of Zn signal is above 50%. The maximum substitution fraction of C is around 12% (calculated by 0.24 $\mu_B$/2.02 $\mu_B$). After considering these two facts the channelling effect for C becomes too weak to be detected by NRA.

In summary, 1 *at.* % to 5 *at.*% C were implanted into the ZnO films on *a*-plane sapphire substrates prepared by PLD. In contrast to the virgin ZnO, the Ne-implanted ZnO and the C-implanted Ge reference samples, clear ferromagnetic hysteresis loops have been observed in all C-implanted ZnO films up to 300 K, indicating a Curie temperature higher than room temperature. The observed ferromagnetism must be related to the chemical involvement of carbon. A saturation magnetic moment of 0.24 $\mu_B$/C at 5 K is observed for the 1 *at.*% C implanted ZnO film and it decreases with increasing carbon concentration. The experiments clearly show that the C-induced ferromagnetism in ZnO can be achieved by different preparation methods. Optimal implantation and post-annealing treatments are needed to improve the effective substitution of C on O sites.

Acknowledgements:

This work is partially (S. Z., Q. X., H. H., and H.S.) supported by BMBF (Grant No. FKZ03N8708 and CHN 05/010). Q. X. acknowledges the National Natural Science Foundation of China (50802041). The authors would like to thank G. Ramm for the PLD target preparation.

Figure captions:

Fig. 1  The *M-H* curves measured at 5 K for virgin, Ne-, and 5 *at.*% C-implanted ZnO with the same sample size of 1 cm$^2$ and mass of 200 mg. The inset displays the M-H curve at 4 K for C-implanted Ge, which exhibits only diamagnetism. Every M-H curve was measured with the field up to 8000 Oe, but only a zoom of the low-field part is shown.

Fig. 2  (a) The *M-H* curves measured at 5 K for 1 *at.*% and 5 *at.*% C-doped ZnO films. The inset shows the *M-H* curve for the 5 *at.*% C-doped ZnO film at 300 K. (b) The ZFC and FC *M-T* curves for 5 *at.*% C-doped ZnO film.

Fig. 3  Carbon-concentration dependent structural and magnetic properties of C-implanted ZnO films. No scaling relation between defects and magnetization can be found.

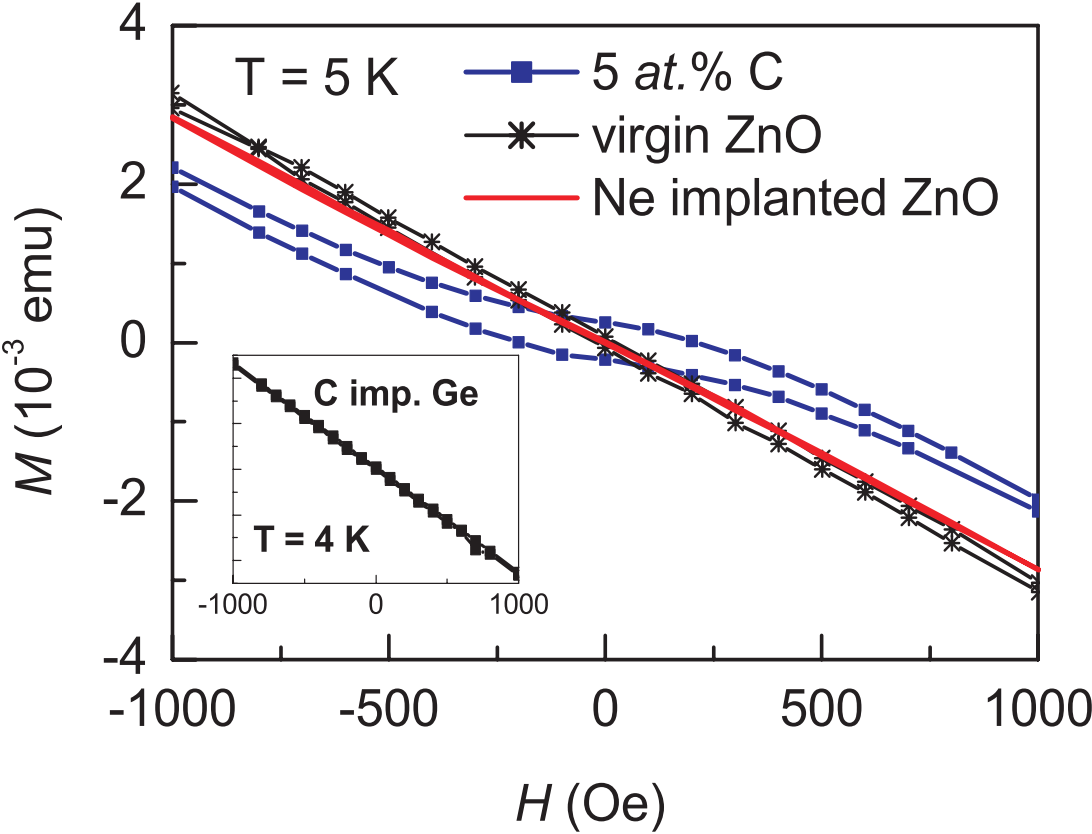

Fig. 1

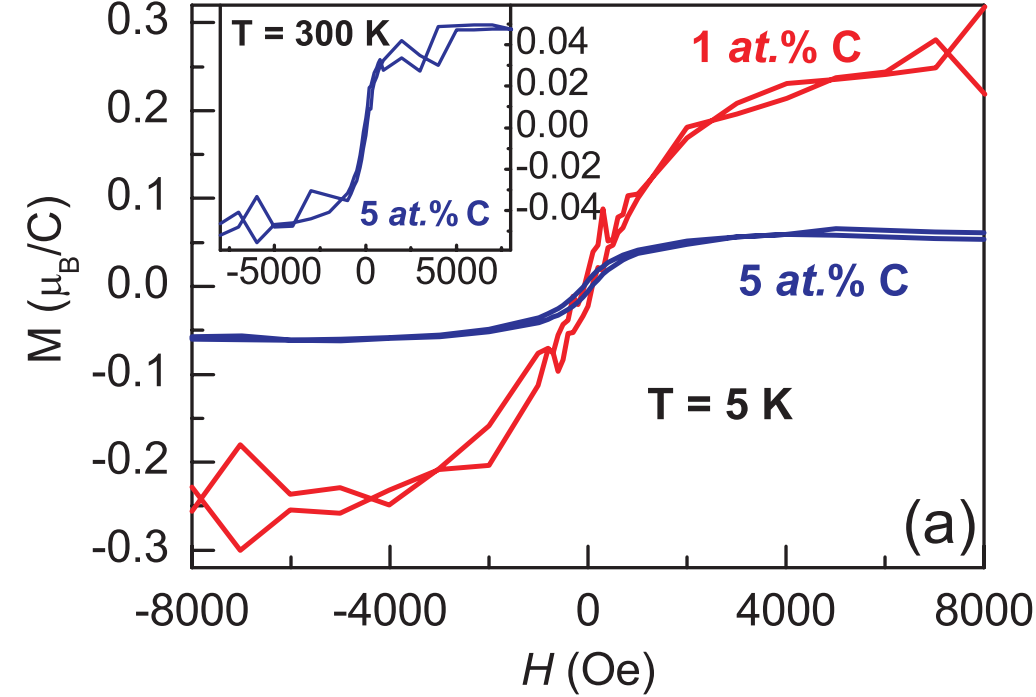
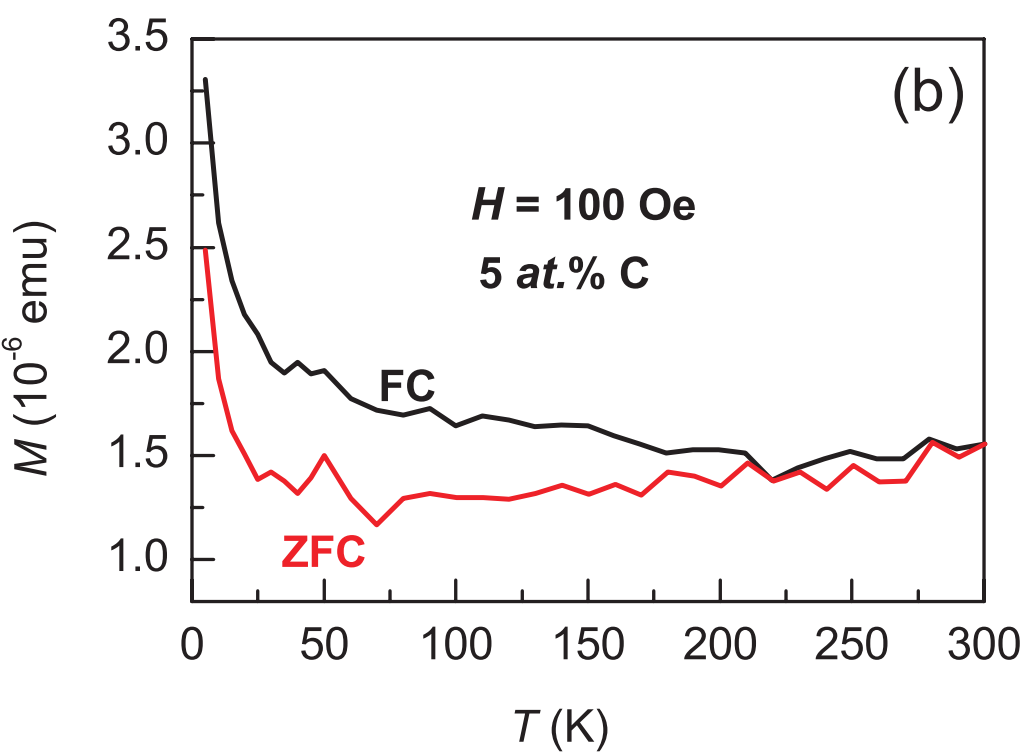

Fig. 2

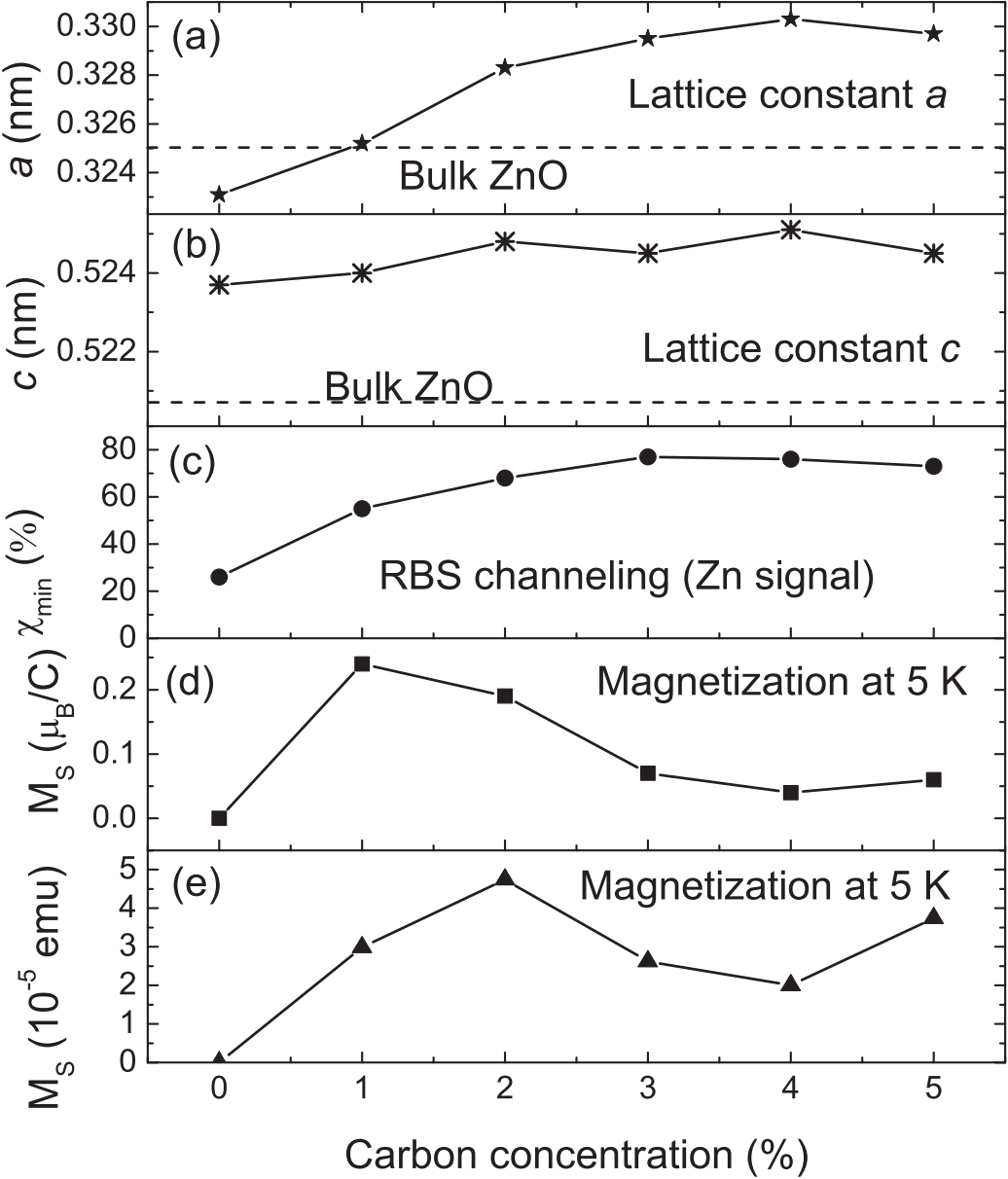

Fig. 3